\documentclass[12pt]{article}

\usepackage[table, dvipsnames]{xcolor}
\usepackage{setspace}
\usepackage{calc,graphics}
\usepackage{changepage}
\usepackage[utf8]{inputenc}
\usepackage{textcomp,marvosym}
\usepackage{fixltx2e}
\usepackage{amsmath,amssymb}
\usepackage{cite}
\usepackage[right]{lineno}
\usepackage{acronym}
\usepackage{times}
\usepackage{bm}
\usepackage{comment}
\usepackage[plain,noend]{algorithm2e}
\usepackage{amsgen,amsmath,amssymb,latexsym}
\usepackage{graphicx}
\usepackage{enumerate}
\usepackage{type1cm}
\usepackage{xspace}
\usepackage{tikz}
\usepackage{float,lscape}
\usepackage{longtable}
\usepackage{makeidx}
\usepackage{multirow}
\usepackage{multicol}
\usepackage{footnote}
\usepackage{rotating}
\usepackage{graphics}
\usepackage{placeins} 
\usepackage{microtype}
\usepackage{geometry}
\usepackage{multirow}
\usepackage{booktabs}
\usepackage{pdfpages}
\usepackage{times}
\usepackage{comment}
\usepackage[plain,noend]{algorithm2e}
\usepackage{arydshln}
\usepackage{amsmath}
\usepackage{type1cm}
\usepackage{tikz}
\usepackage{color}
\usepackage{rotating}
\usepackage{blindtext}
\usepackage{transparent}
\usepackage{setspace}
\DeclareGraphicsExtensions{.PNG, .pdf}

\newcommand{\indep}{\mbox{$\,\perp\!\!\!\perp\,$}}
\newcommand{\nindep}{\mbox{$\,\not\!\perp\!\!\!\perp\,$}}
\newcommand{\q}{``}
\newcommand{\qq}{''\xspace}

\usetikzlibrary{shapes,arrows,shadows,decorations}
\usetikzlibrary{decorations.pathmorphing}
\usetikzlibrary{backgrounds}

\newcommand{\newpar}{{\vspace{0.15cm} \noindent}}
\newcommand{\x}{\noindent\verb&}

\begin{document}

\title{Bayesian Mendelian Randomization for incomplete
pedigree data, and the characterisation of Multiple Sclerosis proteins}

\author{Teresa Fazia$^{1,\dagger\ast}$,
Leonardo Egidi $^{2,\dagger}$,
Burcu Ayoglu $^{3}$,
Ashley Beecham $^{4}$,\\
Pier Paolo Bitti $^{5}$,
Anna Ticca $^{6}$,
Jacob L. McCauley $^{4}$,\\
Peter Nilsson $^{3}$,
Carlo Berzuini $^{7\ddagger}$,
Luisa Bernardinelli $^{1\ddagger}$\\[4pt]
\rule{0cm}{1cm}
\textit{$^{\dagger}$} Joint First Authors,\\
\textit{$^{\ddagger}$} Joint Last Authors\\
\rule{0cm}{1cm}
\textit{$^{\ast}$} Corresponding Author: Teresa Fazia,
Department of Brain and Behavioral\\ Sciences,
Via Bassi 21. University of Pavia, 27100 Pavia, Italy.\\
e-mail: teresa.fazia01@ateneopv.it}

\date{}

\markboth
{T.Fazia, L.Egidi and others}
{Bayesian Mendelian Randomisation Analysis of Family Data}

\maketitle

\vspace{-0.5cm}
\noindent \textit{$^{1}$} Department of Brain and
Behavioural Sciences, University of Pavia, Pavia,
Italy \hfill\\
\textit{$^{2}$} Department of Economics, Business,
Mathematics and Statistics “Bruno de \hfill\\ Finetti” (DEAMS),
University of Trieste, Trieste, Italy \hfill\\
\textit{$^{3}$} SciLifeLab, Dept. of Protein Science,
KTH Royal Institute of Technology,\hfill\\
Stockholm, Sweden\\
\textit{$^{4}$} John P. Hussmann Institute for Human
Genomics and Dr John Macdonald\\ Foundation, Department
of Human Genetics, University of Miami, Miller\\
School of Medicine, Miami, USA\\
\textit{$^{5}$} Azienda Tutela Salute Sardegna.
ASSL Nuoro.
Immunoematologia e Medicina\\
Trasfusionale. Ospedale "San Francesco", Nuoro, Italy\\
\textit{$^{6}$} Azienda Tutela Salute Sardegna.
ASSL Nuoro. Neurologia e Stroke Unit.\\
Ospedale "San Francesco", Nuoro, Italy\\
\textit{$^{7}$} Centre
for Biostatistics,
The University of Manchester,
Jean McFarlane Building,\\
University Place, Oxford Road,
Manchester
M13 9PL, UK\\


\section*{Introduction}

\noindent Before the GWAS (genome-wide association study) 
era, many genetic determinants of disease were 
found via analysis of multiplex pedigrees, that is, by 
looking for genetic markers that run in families in a similar 
way as disease. GWAS advent has robbed pedigree 
analysis of its luster. Future
scientific methodology seesaw might bring pedigree analysis back 
into the spotlight.

\newpar After the recent discovery of hundreds of 
disease-associated variants, interest is focusing on the 
way these variants affect downstream molecular 
markers, such as transcripts and protein levels, and 
on the way the resulting changes in these markers 
in turn affect disease risk. Statistical methods such 
as Mendelian Randomization \cite{Katan1986}, 
hereafter denoted as MR, represent important tools 
in this effort. Most MR studies are based on data from 
unrelated individuals, a notable exception being
\cite{Brumpton2019}. In the present paper we argue 
that by enriching these data with data from 
family-related individuals, a number of difficulties 
that are encountered in MR can be significantly attenuated.

\newpar Motivated by the above considerations, this 
paper discusses extensions of MR to deal with pedigree 
data. We adopt the Bayesian MR framework proposed 
by Berzuini and colleagues \cite{Berzuini2018a}, and 
extend it in various ways to deal with pedigree data. 
The proposed method exploits recent developments 
in Markov chain Monte Carlo (MCMC) inference, as 
offered by the {\tt Stan} probabilistic programming 
language \cite{carpenter2017}.

\newpar We illustrate the method with the aid of data 
generated by ImmunoChip genotyping and 
transcriptome/protein assays on members of Multiple 
Sclerosis (MS) multiplex pedigrees from an isolated 
Sardinian (italian island) population. With this kind 
of data, environmental  confounding and population 
stratification are expected to have less impact on causal 
effect estimates, and the effects of rare variants to be 
easier to detect. Thanks to our Bayesian technology, 
we perform a "clever" analysis where an initial model 
is gradually elaborated to bring biological theory and 
relevant information in general to bear. In this 
paper, we include in the MR model such information as 
a family indicator, parental protein levels and kinship. 
Not only do such enhancements provide extra protection 
against bias, but they also allow us to explore a number 
of secondary aspects of the biological mechanism. A 
further advantage of the Bayesian approach is the 
simple way it deals with incomplete information. In 
our study, missing values of the exposure (the level of 
a protein) are treated as additional parameters to be 
estimated from the data, without incurring biases, as 
is natural in Bayesian analysis.

\newpar The "outcome" variable of our analysis is the 
MS disease indicator. MS lends itself well to a MR study. 
This disease tends to become manifest early during 
reproductive lifespan of most humans, throughout 
history, and is therefore likely to have a strong genetic 
component. Genetic variants are therefore expected to 
act as good instruments for the MR analysis. The main 
scientific question in this paper is whether the plasma 
level of IL12A protein (which in our analysis will be 
referred to as the "exposure”)  is causal with respect to 
development of MS (outcome). It is believed that 
dysregulation of circulating proteins is a causal 
determinant in many pathologies, more directly so 
than genetic variants. Our analysis is further motivated 
by the importance of proteins as natural drug targets.  
We could have harnessed publicly available eQTL information to 
involve in the analysis protein concentrations in tissues 
other than blood, but we do not pursue this here, not 
to obscure the main points of the paper, whose main 
message is methodological.

\vspace{1cm}
\section*{Methods}

\subsection*{\Large Sample Description}

\noindent Our MS patients were ascertained through the case
register established in 1995 in the province of Nuoro, Sardinia,
Italy. Cases were diagnosed according to Poser’s 
criteria \cite{Poser1983}. Twenty extended MS multiplex 
pedigrees were selected for the analysis, for a total 
of $N=936$ individuals (98 cases and 838 unaffected relatives). 
A subset of the pedigree members had complete data, consisting 
of the observed levels of the IL12A protein (the exposure), the 
known disease indicator (the outcome variable), and the genotypes 
at all loci of Immunochip (see below). The remaining individuals 
had complete data except for a missing value for the protein level.

\subsection*{\Large Genotyping Data}

\noindent Genotyping data were obtained by using Immunochip 
Illumina Infinium HD custom array (hereafter “Immunochip” 
for brevity), designed for fine mapping of 184 established 
autoimmune loci \cite{Beecham2013}.

\newpar The quality control-filtered dataset included 127134 
Single Nucleotide Polymorphisms (SNPs) across Immunochip
\cite{Fazia2017}. For a first stage of our analysis, we imposed 
a maximum correlation of $r^2 = 0.20$ between candidate 
instrumental SNPs within a 100 Kb window, by using 
the {\tt indep-pairwise} command of the PLINK 
package \cite{Purcell2007}. This yielded a total of 19121 
candidate SNP instruments across Immunochip.

\subsection*{\Large Protein Selection and Profiling}

\noindent The protein we chose for our illustrative study was IL12A.
Choice was made prior to considering the data, on the basis of 
Genome-Wide Significant (GWS) association between MS and genetic 
variants located within (e.g. exonic, intronic, in the UTR) or in 
the proximity (e.g. downstream, intergenic) of the protein-coding 
gene \cite{Beecham2013} and on the basis of literature evidence on 
the biological role of this cytokine in the context of MS \cite{constantinescu1998antibodies} 
 \cite{jahanbani2019serum} \cite{rentzos2008effect} \cite{sun2015interleukin}.
Detailed information about the locations 
of the strongest MS association signals within or in the proximity 
of the protein-coding genes, and about the strengths of the MS 
associations, are reported for IL12A in the Supplementary Material.

\newpar Plasma profiles were analysed by using a bead-based
antibody array format, consisting of polyclonal Human Protein Atlas
\cite{Nilsson2005} 
 antibodies immobilized onto microspheres in suspension
\cite{Schwenk2007} \cite{Schwenk2008} (see Supplementary Material for details).

\subsection*{\Large Selection of Instrumental Variants}
\noindent Genetic variants with a significant marginal 
association ($p<5 \times 10^{-3}$) with the level of the 
protein of interest and mutual  $r^2<0.20$ correlation were 
selected to act as instrumental variables (IVs) in the first 
stage of our analysis. The liberal $p<5 \times 10^{-3}$ threshold 
is justified by the fine genotyping of candidate gene 
regions and by recent arguments
\cite{Wang2016} \cite{Yang2011}
in favour of using sub-genome-wide-significance loci to 
strengthen biologically interesting signals. It is also justified 
by the relative ability of our Bayesian MR method (when compared 
with most frequentist approaches) to deal with the weak instrument 
bias, thanks to the uncertainty of the estimated exposure 
coefficients being explicitly included in the model.

\subsection*{\Large Notation}

\noindent In our analysis, the putative causal factor (with 
respect to disease) is the circulating level of protein IL12A. 
We call this variable the "exposure", and denote it as $X$. 
We let the symbol $\Sigma_X$  denote a regime 
indicator \cite{Dawid2000} \cite{Dawid2002} which tells us 
whether we are considering the {\em actual} data generating 
regime for $X$, which is observational, or a {\em hypothetical} 
regime where variable $X$ in each individual is set to a 
value $x$ by intervention. The observational regime corresponds 
to $\Sigma_X=\emptyset$, whereas the latter, interventional, 
regime corresponds to $\Sigma_X=x$. In our analysis the outcome 
variable, $Y$, indicates whether the individual has the 
disease ($Y=1$) or not ($Y=0$). We are interested in the "causal 
effect" of $X$ on $Y$, that is, in the way the distribution 
of $Y$ changes when $X$ is first set by intervention to a 
reference value $x_0$ and then forced to take the new 
value $x_1$. Throughout this paper we take this causal effect 
to be defined as the causal odds ratio (COR):
\begin{equation}
COR = \frac{P(Y=1\mid \Sigma_X = x_1)} {P(Y=1\mid \Sigma_X=x_0)}
 \frac{1-P(Y=1\mid \Sigma_X = x_0)} {1-P(Y=1\mid \Sigma_X=x_1)} 
\end{equation}
\noindent  The reason why we can't generally measure causal 
effect by standard regression of $Y$ on $X$ is that the regression 
coefficient will have no causal interpretation in the presence of 
unobserved confounders of the exposure-outcome relationship, 
which we denote as $U$. This is, indeed, why we need to use MR. 
We shall model $U$ as an individual-level scalar variable, more 
precisely, a one-dimensional reduction of the unknown collection 
of confounders. MR requires availability of a set of instrumental 
variables,  or instruments, denoted 
as $Z  \equiv (Z_1, \ldots , Z_J) $, which in a standard analysis 
will often correspond to the individual's genotypes at a set of SNP 
loci. Each of these genotypes we code as "allele doses", with 
values $(0,1,2)$ respectively indicating presence of zero, one and 
two copies of the "alternative" allele at the locus. For most 
individuals in the pedigree, we also have observed {\em (i)} 
maternal and paternal genotypes at each instrumental locus 
and {\em (ii)} the levels of protein IL12A in the father and 
in the mother. Let the collection of maternal (rep., paternal) 
genotypes for the generic individual be denoted as $Z_m$ ($Z_p$). 
Let the protein levels for the mother and the father of the generic 
individual be denoted as $W_M$and $W_F$, respectively . We further 
introduce an individual-level categorical variable, denoted as $F$, 
which indicates the individual's pedigree of membership, or family. 
Further notation will be introduced in the next sections, as required.

\subsection*{\Large Assumptions}

\noindent  This paper uses Dawid's conditional independence formalism
\cite{Dawid1979}, with the $\indep$ symbol
representing conditional independence, so that $A \indep B \mid C$, stands 
for \q$A$ is independent of $B$ given $C$, and $A \nindep B$, means \q$A$ 
is not independent of $B$ \qq. Conditions introduced in this section are
required for method validity, except for one of them. They are essentially
identical to those required by standard MR  methods.

\newpar Here are the assumptions. Each $j$th instrumental variable, $Z_j$,
must satisfy the {\em confounder independence} condition $Z_j \indep U$, 
stating that the instrument is unrelated to exposure-outcome confounders. 
A further condition called {\em exclusion-restriction} requires that
 $Y \indep Z_j \mid (X,U)$, that is,  each $j$th instrument can be 
associated with response only via the exposure. Exclusion-restriction 
is a desirable condition, however, unlike the remaining conditions in 
this section, it is not required by our method. Next comes the {\em 
instrument relevance} condition, $Z_j \nindep X$, stating that no 
instrument is independent of the exposure. We have also conditions 
involving the regime indicator, $\Sigma_X$.  The {\em confounder 
invariance} condition, $U \indep \Sigma_X$, requires that the 
distribution of the confounders $U$ be the same, whether or not we 
intervene on $X$, and regardless of the value imposed on or observed 
in $X$. Next comes the {\em interventional irrelevance} condition
$\Sigma_X \indep Z$, requiring that any intervention on $X$ has no 
consequence on $Z$, and the {\em interventional modularity} 
condition, $\Sigma_X \indep Y \mid (X,U)$, asserting that once 
we are told the values of $X$ and $U$, the distribution of $Y$ no 
longer depends on the way the value of $X$ has arisen, whether 
observationally or through the intervention of interest.

\newpar Those independence relationships that involve the (non-stochastic) 
regime indicator should be interpreted in the light of the extended 
conditional independence calculus described by Constantinou et al 
\cite{constantinou2017}. The relationships between $\Sigma_X$ and 
the remaining variables, as depicted in Figure 1, characterize the 
influence of $X$ on $Y$, corresponding to the $X \rightarrow Y$ 
arrow, as causal. The remaining arrows in the graph, eg $Z \rightarrow X$, 
do not necessarily have to be interpreted as causal, which greatly 
expands method applicability.

\newpar How realistic are the above assumptions? This is a crucial 
question, considering that all the above assumptions, except for 
instrumental relevance, are at best only indirectly testable, or
corroborated on the  basis of bological knowledge. Take, for example, 
the confounder independence assumption. In our application, where 
the exposure is a low-level biological mark, it may be reasonable to 
assume that those genetic variants that operate in {\em cis} with 
respect to the studied protein, exert no effect on common causal 
precursors of exposure and outcome other than effects mediated 
by the exposure. This assumption can be further corroborated 
by investigations based on eQTL data and on the known linkage 
disequilibrium (LD) pattern in the DNA region of interest. The 
assumption of confounder invariance requires more attention than 
is usually the case. In our application, for example, if the intervention 
represented by $\Sigma_X$ consisted of a particular diet, then 
confounder invariance would be violated, because a diet will hardly 
modify the level of the protein without altering a constellation 
of metabolites that act as potential confounders. Interventional 
irrelevance is defendable in our applicative situation, by using 
randomization arguments. As concerns interventional modularity, 
in our study this condition implies, in particular, that a unit increase 
in $X$ caused by one of the variants in the instrumental set should 
exert on $Y$ the same effect as a unit increase in $X$ caused by the 
intervention of interest. In our application, where the instrumental 
effects are regulatory and the intervention of interest consists of a 
pharmacological modification of $X$, interventional modularity 
appears to be a defendable assumption.

\newpar All the conditions defined above, except for exclusion-restriction, 
are required by our method.

\newpar Sometimes it is possible, and then helpful, to represent
the qualitative structure of a statistical model by a directed acyclic 
graph \cite{Lauritzen1996}. A stripped-down representation of the 
class of MR models discussed in the present paper
is shown in Figure 1. All the
conditions stated above (except for exclusion-restriction)
can be read off the graph of Figure 1 by applying 
$d$-separation \cite{Geiger1990} or moralization \cite{Lauritzen1996},
with the following additional rules:  {\em (i)} faithfulness \cite{Spirtes2001} 
of the $Z \rightarrow X$  edges (which means assuming that any 
distribution which follows the 
model only exhibits independence relations represented by the directed 
acyclic graph), and  {\em (ii)} assigning a value $x$ to $\Sigma_X$ implies 
the simultaneous assignment of the same value to $X$, and  {\em (iii)}  
assigning a value $x$ to $\Sigma_X$ implies that all arrows into $X$ 
except for $\Sigma_X \rightarrow X$ are severed. Because most of the 
conditions introduced at the beginning of this section are not directly 
testable on the basis of the data, the Reader should be aware that 
graphs like the one shown in Figure 1 describe an {\em assumed}, 
ultimately uncertified, albeit plausible, state of affairs. We shall assume 
throughout the paper that the above described conditions, bar 
exclusion-restriction, are valid. 

\newpar We conclude this section with a brief discussion of the 
exclusion-restriction assumption. This assumption (which is not 
required by our method) does not allow an instrument to exert 
an effect on $Y$ other than that exerted though the mediating 
effect of $X$. In our graph of Figure 1, this condition is violated 
by the $Z_J \rightarrow Y$ arrow. Because of this, the effect of 
instrument $Z_J$ on $Y$ is said to be \q pleiotropic \qq according 
to Figure 1. In the context of our application, pleiotropic effects 
may arise from two broad classes of mechanism. The first is due 
to the eQTL variants used as instruments being in linkage 
disequilibrium (LD) with eQTLs of nearby genes. The second is 
due to the instrumental variant exerting a causal effect on $Y$ 
through a pathway independent of $X$. Although the former type 
of pleiotropy could, in principle, be neutralized by conditioning on 
the eQTLs in the region, except for the instrumental variants, the 
latter cannot be directly tested from the data. It would therefore be 
uncautious to perform MR by using a method that does {\em not} 
allow for general types of pleiotropy. Our Bayesian approach deals 
with the problem by explicitly introducing the unknown pleiotropic 
effects in the model, and by treating them as unknown parameters 
to be estimated from the data.


\begin{center}
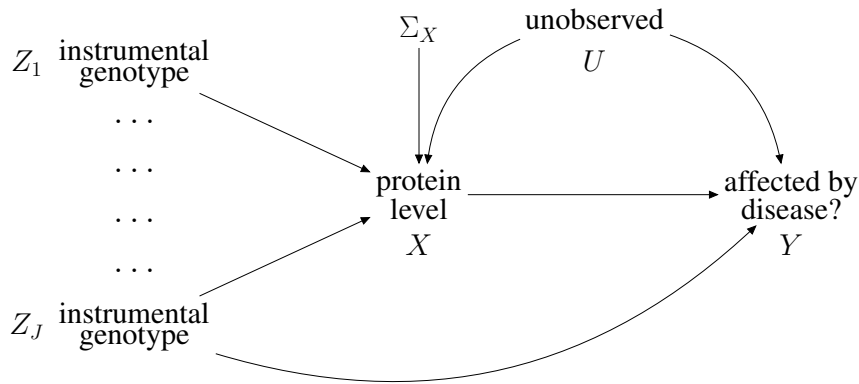
\begin{figure}
\centering
\caption{\footnotesize Graphical representation of a Mendelian
randomization model for the analysis of unrelated  individuals.}
\scalebox{0.55}{
\begin{tikzpicture}[align=center,node distance=4.5cm]
\node (Uchild) [fill=none,circle] {\LARGE unobserved};
\node (Ulab) [below of=Uchild,fill=none,node distance=1cm]
{\LARGE $U$};
\node (Xchild) [below left of=Uchild,fill=none,rectangle,node distance=6cm]
{\LARGE  \LARGE protein \\ \LARGE level};
\node (Xlabchild) [below of=Xchild,fill=none,node distance=1.2cm]
{\LARGE $X$};
\node (Ychild) [right of=Xchild,fill=none,
rectangle,node distance=9cm]  {\LARGE affected by\\ \LARGE disease?};
\node (Ylabchild) [below of=Ychild,fill=none,node distance=1
.2cm] {\LARGE $Y$};
\node (handle) [left of=Xchild,fill=none,node distance=3.7cm]{};
\node (ZchildJ) [below left of=handle,fill=none,rectangle]
{\LARGE  \LARGE instrumental \\ \LARGE genotype};
\node (ZlabchildJ) [left of=ZchildJ,fill=none,node distance=2.6cm] {\LARGE $Z_J$};
\node (Zchild1) [above left of=handle,fill=none,rectangle]
{\LARGE  \LARGE instrumental \\ \LARGE genotype};
\node (Zlabchild1) [left of=Zchild1,fill=none,node distance=2.6cm] {\LARGE $Z_1$};
\node (dot1) [below of=Zchild1,fill=none,node distance=1.4cm] {\huge $\ldots$};
\node (dot2) [below of=dot1,fill=none,node distance=1.2cm] {\huge $\ldots$};
\node (dot3) [below of=dot2,fill=none,node distance=1.2cm] {\huge $\ldots$};
\node (dot4) [below of=dot3,fill=none,node distance=1.2cm] {\huge $\ldots$};
\node (F) [above of=Xchild,fill=none,node distance=4cm]
{\LARGE $\Sigma_X$};
\draw[-triangle 45,bend left] (Uchild) to node {} (Ychild);
\draw[-triangle 45, bend right] (Uchild) to node {} (Xchild);
\draw[-triangle 45] (Zchild1) to node {} (Xchild);
\draw[-triangle 45] (ZchildJ) to node {} (Xchild);
\draw[-triangle 45] (Xchild) to node {} (Ychild);
\draw[-triangle 45] (F) to node {} (Xchild);
\draw[-triangle 45, bend right] (ZchildJ) to node [above] {} (Ychild);
\end{tikzpicture}
}
\end{figure}
\end{center}

\subsection*{\Large Progressive Elaboration of the Model}

\noindent A "naive" approach consists of analyzing the 
pedigree data by using the Bayesian MR model proposed by 
Berzuini and colleagues, as described in \cite{Berzuini2018a}, 
as if the individuals were independent. This will, of course, 
produce biased estimates. We shall use this "independence model" 
in a preliminary analysis of the data. We shall then step through 
a sequence of re-analyses of the data based on more elaborated, 
and more realistic, models, that  we describe in the following.

\vspace{2.5cm}
\begin{center}
{\large Independence Model}
\end{center}
\vspace{0.1cm}

\newpar The model of Berzuini and colleagues
\cite{Berzuini2018a} assumes that individuals are independent,
and that the $X$ variable has been standardized to 
have zero mean and unit standard deviation. The data 
generating equations of the model conform with the 
conditional independence assumptions expressed in 
Figure 1, and take the form:
\begin{eqnarray}
\label{full1}
P(U) &=& \mbox{N}(0,1),\\
\label{full2}
P(X \mid Z_1, \ldots , Z_J, U) &=& \mbox{N}(
\sum_{j=1}^J \alpha_j Z_j +
\delta_X U, \sigma_X^2),\\
\label{full3}
P(Y \mid X, Z_1, \ldots , Z_J, U) &=& \mbox{logit}^{-1}(\omega_Y + \theta X +
\sum_{j=1}^J \beta_j Z_j + U),
\end{eqnarray}
\noindent where $\mbox{N}(a,b)$ stands for a
normal distribution with
mean $a$ and variance $b$, the symbol
$\alpha
\equiv (\alpha_1, \ldots , \alpha_J)$
denotes the instrument (i)-exposure (e) associations and
$\beta
\equiv (\beta_1, \ldots , \beta_J)$
are the pleiotropic effects. The only difference from 
Berzuini et al here is that the outcome variable $Y$ is 
no longer normal, but Bernoulli, as appropriate for a 
binary random variable. Recall that, in our study, some 
components of the $X$ vector (protein level measurements) 
are missing, which is not made explicit in the notation. 
The Bayesian inference engine identifies the missing components 
and treats them as unknown parameters, effectively integrating 
them out to obtain the posterior distribution for the 
parameters of inferential interest. Note that this way 
of dealing with missing data is more efficient than, say, 
imputing each missing component of $X$ on the basis of the 
individual's observed $Z$ values, thanks to the fact that, 
in our method, the missing values are estimated by using 
information about both $X$ and $Y$.

\newpar In the above equations, the causal effect of interest,
denoted as $\theta$,
represents the change in log-odds of probability of $Y=1$
caused by an interventional change of
one standard deviation in $X$.

\newpar As shown in \cite{Berzuini2018a} for the normal case,
parameters  $(\alpha,
\tau_X)$ are identified by the data, but
the remaining parameters, including
the causal effect of interest, $\theta$,
are not. Berzuini  and colleagues deal with the problem 
by a combination of two devices. The first consists of 
introducing the additional (untestable) assumption that 
each $j$th component of $\beta$ is a priori independent 
of the remaining parameters of the model, formally,
$P(\beta_j \mid \alpha_j,\tau_X) = P(\beta_j)$.
This is called the Instrument Effects Orthogonality (IEO) condition.
The second consists of introducing
a proper, scientifically plausible, prior for $\beta$, which 
makes inferences possible by inducing on $\theta$ (and on further 
parameters of potential posterior interest) a proper posterior. 

\noindent As concerns the prior component of our Bayesian model, 
we invite the Reader to consult \cite{Berzuini2018a}.

\newpar Variations have been introduced. While still imposing on the
pleiotropic effects $\beta$ a horseshoe prior \cite{carvalho2010}, 
we are now using the enhanced version of this distribution 
proposed by Piironen and Vehtari \cite{Piironen2017}. Also, we 
take $\theta$ -- the causal effect of main inferential 
interest -- to have a Cauchy(0,2.5) prior, with the following 
justification. Because $X$ has been standardized to have mean 
0 and unit standard deviation (SD), the mentioned prior 
for $\theta$ states as unlikely that a one-SD change in 
protein level causes a change in risk of disease 
exceeding 5 points on a logit scale, which corresponds to 
shifting a probability of disease  occurrence from, say, 
0.01, to 0.5, or from 0.5 to 0.99. This is also in 
agreement with current evidence on the effect of 
circulating proteins on disease  \cite{Sun2018}.

\newpar Finally, we are now taking the i-e 
associations, $\alpha$, to be independently distributed 
according to a double-exponential distribution with mean 0 
and unknown scale. One merit of this prior is to shrink the 
small effects to zero, which reduces the weak instrument bias, 
so that the model works with an adaptively selected subset 
of strong instruments.

\vspace{0.3cm}
\begin{center}
{\large Introducing Kinship}
\end{center}

\noindent Treating members of a pedigree as independent
individuals, which they are not, will produce overconfident
and biased estimates. We remedy this by introducing in the
model between-individual correlation in the form of the
kinship matrix, which can be derived by a standard algorithm
from the structure of the pedigree. We are currently working
with a single, overarching, kinship matrix of size $N \times N$,
where $N$ is the total number of individuals in the sample. This
large matrix contains zeros corresponding to pairs of individuals
in different families. The method could be made computationally
more efficient by introducing  family-specific matrices.
Kinship information is introduced in the model by writing:

\begin{eqnarray}
\label{kinship1}
\label{full3}
P(Y \mid X, Z_1, \ldots , Z_J, U) &=& \mbox{Bernoulli}(\pi),\\
logit(\pi) &=& \mbox{MVN}(\mu, \Sigma),\\
\mu &=& \omega_Y + \theta X +
\sum_{j=1}^J \beta_j Z_j + U,
\end{eqnarray}
\noindent where $\Sigma$ is the $N \times N$ kinship matrix,
the notation $\mbox{MVN}   (a,b)$ stands for 
multivariate normal distribution with vector mean $a$ 
and variance-covariance matrix $b$.

\vspace{0.3cm}
\begin{center}
{\large Introducing Family Effects}
\end{center} 

\newpar In our analysis, we incorporate family information 
simply by designating a categorical variable $F$ to indicate 
the individual's family, with $F \in (1, \ldots, M)$, with 
$M=12$, and by modifying the outcome and exposure 
models to take the following form:

\begin{eqnarray*}
\label{family1}
P(X \mid Z_1, \ldots , Z_J, U, F) &=& \mbox{N}(\nu,\sigma_X^2),\\
\nu &=& \sum_{j=1}^J \alpha_j Z_j +
\delta_X U+
\sum_{f=1}^{M} I_{F=f} \; \gamma^X_f ,\\
\label{family2}
P(Y \mid X, Z_1, \ldots , Z_J, U, F) &=& {\rm Bernoulli}(\pi),\\
logit(\pi) &=& \mbox{MVN}(\mu, \Sigma),\\
\mu &=& \omega_Y + \theta X +
\sum_{j=1}^J \beta_j Z_j + U+
\sum_{f=1}^{M} I_{F=f}\; \gamma^Y_f,
\end{eqnarray*}
\noindent where $I_A$ stands for the indicator function, taking 
value 1 if the logical condition $A$ is true, and value 0 otherwise. 
The quantities $\gamma^X \equiv (\gamma^X_1, \ldots , 
\gamma^X_M)$ and $\gamma^Y \equiv (\gamma^Y_1, 
\ldots , \gamma^Y_M)$ are vectors of unknown "family 
effects", respectively on $X$ and on $Y$. In our analysis, we 
have imposed on these parameters independent and mildly 
informative priors, with greater spread than the prior 
for $\theta$.

\newpar The family indicator appears in the graph of Figure 2
with the symbol $F$. According to this graph, failure to condition on
this indicator (that is, removing the $F$ variable from the 
model) "opens" (unblocks) the $Z \leftarrow F \rightarrow Y$ path, 
and the $Z \leftarrow F \rightarrow U$ path, in the terminology 
of $d$-separation. Which means that failure to condition on 
family creates  a spurious, exposure-unmediated, association 
between instrument and outcome and, what's even worse, 
violates the Confounder Independence assumptions. Hence, 
inclusion of the family indicator in the model prevents the 
estimate of the causal effect from being unduly distorted. 
In situations where the sample contains unrelated (in 
addition to related) individuals, the unrelateds may be 
lumped into a single, notional, family.

\begin{center}
\begin{figure}
\centering
\caption{\footnotesize Incorporating a family indicator variable ($F$).}
\scalebox{0.55}{
\begin{tikzpicture}[align=center,node distance=4.5cm]
\node (Uchild) [fill=none] {\LARGE unobserved};
\node (Ulab) [below of=Uchild,fill=none,node distance=1cm]
{\LARGE $U$};
\node (family) [above of=Uchild, fill=none,node distance=4cm]
{\LARGE Family\\ \LARGE indicator};
\node (Flab) [above of=family,fill=none,node distance=1cm]
{\LARGE $F$};
\node (Xchild) [below left of=Uchild,fill=none,rectangle,node distance=6cm]
{\LARGE  \LARGE protein \\ \LARGE level};
\node (Xlabchild) [below of=Xchild,fill=none,node distance=1.2cm]
{\LARGE $X$};
\node (Ychild) [right of=Xchild,fill=none,node distance=9cm]
{\LARGE affected by\\ \LARGE disease?};
\node (Ylabchild) [below of=Ychild,fill=none,node distance=1
.2cm] {\LARGE $Y$};
\node (handle) [left of=Xchild,fill=none,node distance=3.7cm]{};
\node (Zchild) [left of=handle,fill=none,rectangle]
{\LARGE  \LARGE instrumental \\ \LARGE genotypes};
\node (Zlabchild) [below of=Zchild,fill=none,node distance=1.2cm] {\LARGE $Z$};
\node (Zlabchild) [below of=Zchild,fill=none,node distance=1.2cm] {\LARGE $Z$};
\draw[-triangle 45, bend left] (Zchild) to node {} (family);
\draw[-triangle 45,bend left] (Uchild) to node {} (Ychild);
\draw[-triangle 45, bend right] (Uchild) to node {} (Xchild);
\draw[-triangle 45] (Zchild) to node {} (Xchild);
\draw[-triangle 45] (Xchild) to node {} (Ychild);
\draw[-triangle 45] (family) to node {} (Uchild);
\draw[-triangle 45,bend right] (family) to node {} (Xchild);
\draw[-triangle 45, bend left] (family) to node {} (Ychild);
\draw[-triangle 45, bend right] (Zchild) to node [above] {} (Ychild);
\end{tikzpicture}
}
\end{figure}
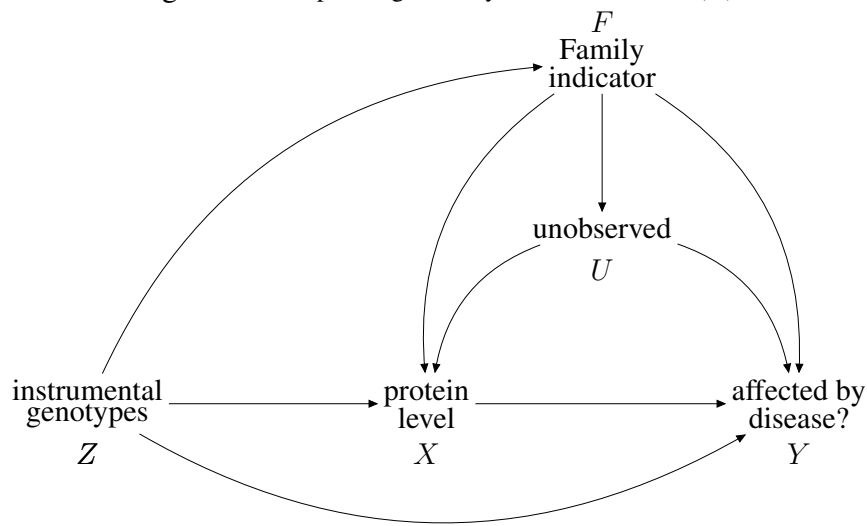
\end{center}

\vspace{0.3cm}
\begin{center}
{\large Introducing Parental Protein Information}
\end{center}

\noindent In this final elaboration step of the model 
we introduce information about the measured level of 
protein in the individual's parents. This is motivated 
by the assumption that there are unobserved loci in DNA, 
denoted by $Z'$,  that (individully or collectively) have 
an effect on the protein of interest. The individual's 
protein level becomes associated with that of their parents 
through $Z'$. And, because of this, parental protein level 
become additional candidate instruments in the analysis. 
We incorporate parental protein information simply by 
designating the continuous variables $P_M$ and $P_F$ to 
represent the measured level of circulating IL12A protein 
in the individual's mother and father, respectively, after 
standardizing them to have zero mean and unit variance. 
The two variables are incorporated in the exposure 
model by writing:

\begin{eqnarray*}
\label{protein}
P(X \mid Z_1, \ldots , Z_J, U, F, P_M, P_F) &=& \mbox{N}(\nu,\sigma_X^2),\\
\nu &=& \sum_{j=1}^J \alpha_j Z_j +
\delta_X U+
\sum_{f=1}^{M} I_{F=f} \; \gamma^X_f  + \alpha^M P_M \alpha^F P_F
\end{eqnarray*}
\noindent with $\alpha^M$ and $\alpha^F$ to be estimated from
the data. It can be shown (but this is outside the scope of the present
work) that the modification is valid provided we assume that $Z$ and $Z'$
are not correlated, and that $Z'$ does not influence $Y$ other than through changes in $X$.

\section*{Results}

 \vspace{0.1cm}
\begin{center}
{\Large Results from Initial Model}
\end{center}

\noindent Estimates of the causal effect of the circulating 
level of IL12A on risk of MS were obtained by using {\tt R} 
package {\tt MendelianRandomization} \cite{yavorska2017}, as 
found on  http://cran.r-project.org. The frequentist causal 
effect estimates, expressed on a log-odds-ratio scale with 
their corresponding 95$\%$ confidence intervals, are summarised 
in Table 1. Difficulties introduced by the missing IL12A values 
have been sidestepped in the simplest way:  by discarding 
individuals who had a missing IL12A value when 
calculating the i-e associations.

\begin{table}[ht]
\centering
\begin{tabular}{rlrrrrr}
\hline
& Method & Estimate & Std Error & 
 \multicolumn{2}{c}{95\% confidence interval}& P-value \\ 
\hline
1 & Simple median & -0.30 & 0.15 & -0.59 & -0.02 & 0.04 \\ 
2 & Weighted median & -0.07 & 0.14 & -0.34 & 0.20 & 0.61 \\ 
3 & Penalized weighted median & -0.15 & 0.14 & -0.42 & 0.12 & 0.28 \\ 
4 & IVW & -0.14 & 0.09 & -0.33 & 0.04 & 0.12 \\ 
5 & Penalized IVW & -0.21 & 0.10 & -0.40 & -0.02 & 0.03 \\ 
 6 & Robust IVW & -0.21 & 0.12 & -0.44 & 0.02 & 0.08 \\ 
7 & Penalized robust IVW & -0.23 & 0.10 & -0.42 & -0.04 & 0.02 \\ 
8 & MR-Egger & 0.51 & 0.37 & -0.22 & 1.25 & 0.17 \\ 
9 & Penalized MR-Egger & 0.51 & 0.37 & -0.22 & 1.25 & 0.17 \\ 
10 & Robust MR-Egger & 0.51 & 1.01 & -1.48 & 2.50 & 0.61 \\ 
11 & Penalized robust MR-Egger & 0.51 & 1.01 & -1.48 & 2.50 & 0.61 \\ 
\hline
\end{tabular}
\label{Table 1}
\caption{\footnotesize Estimates for the causal effect of the 
circulating level of IL12A on risk of MS obtained by 
using {\tt R} package Mendelian ({\tt http://cran.r-project.org}). 
Estimated causal effects are expressed on a log-odds-ratio scale. }
\end{table}

\newpar According to Table 1, estimates from the frequentist 
MR methods considered in this paper exhibit a poor consistency. 
A significant estimate of the causal effect was obtained only with 
the Simple Median and with the Penalized IVW methods, the latter 
requiring the assumption of no pleiotropy.

\newpar The model by Berzuini and colleagues \cite{Berzuini2018a}, 
which also assumes sample individuals to be independent of each 
other (see Methods section), gave an estimated log-odds-ratio 
causal effect of -0.202, with a standard error of 0.078, and a
95$\%$ credible interval of -0.418 through -0.091. This result 
was obtained by treating the missing protein levels as additional 
unknown parameters to be estimated from the data.

\vspace{2cm}

\begin{center}
{\Large Results after Introducing Kinship} 
\end{center}

\newpar Our frequentist analyses were repeated in a sounder 
fashion, by estimating the disease-instrument log-odds-ratio
associations via a mixed-effects model  ({\tt lmekin} 
function of  {\tt R}, as described in \cite{Pinheiro2000}), 
that allows family relationships between pedigree members, 
as expressed by the kinship matrix, to be taken into 
account \cite{Fazia2017}.  Significant estimates were 
then obtained by using IVW ($\hat{\theta}=-0.18, p < 0.0001$) 
and WME  ($\hat{\theta}=-0.11, p = 0.012$), but not by using 
MR-ER ($\hat{\theta}=-0.23, p = 0.7$).

\newpar By contrast, when we extended the model by Berzuini 
and colleagues to incorporate family relationships, as 
expressed by the kinship matrix (see Methods section), and used 
it to re-analyse the data, the estimated causal effect was no 
longer significant, as reported in Table 2. This was not 
unexpected, when one considers that between-individual 
correlation reduces the "effective" sample size, and, 
as a consequence, statistical power.

\newcommand{\ra}[1]{\renewcommand{\arraystretch}{#1}}

\vspace{1cm}
\begin{table*}[ht]
\centering
\ra{1.3}
\begin{tabular}{lrrrrr}
\rowcolor{lightgray}
\vspace{-0.1cm}
&\multicolumn{5}{c}{PERCENTILES OF POSTERIOR}\\
\rowcolor{lightgray}
CAUSAL EFFECT OF 1SD CHANGE& \multicolumn{5}{c}{DISTRIBUTION}\\
\rowcolor{lightgray}
IN PROTEIN LEVEL ON MS RISK& $5$ & $25$ & $50$ & $75$ & $95$\\
\midrule
Causal Log Odds Ratio Effect& -0.91 & -0.59 & -0.39 & -0.17 & 0.10  \\ 
Causal Odds Ratio Effect& 0.4 & 0.55 & 0.67 & 0.84 & 1.1  \\
\end{tabular}
\caption{\footnotesize Estimates for the causal effect 
of the circulating level of IL12A on risk of MS obtained 
by using an extension of the model by Berzuini and 
colleagues which incorporates family relationships, as 
expressed by the kinship matrix.}
\end{table*}

\vspace{2.9cm}
\begin{center}
{\Large Results after Introducing Family Effects}
\end{center}

\noindent In the Methods section we have seen that(pedigree) 
membership may introduce bias in the estimated causal effect 
by acting as a confounder of the relationship between 
instrumental genotypes and outcome, in a way similar to 
what population stratification does. This is a consequence 
of the family variable being generally associated with both 
the individual's genetic set-up and with disease-linked 
unobserved factors (genetic variants, environment, education, 
and so on). See the Methods section for a more rigorous 
discussion of the issue. When we introduced both kinship 
information and the family variable (as a 12-level categorical 
factor) in the model, we got the causal effect estimate 
summarised in Table 3.

\vspace{1cm}
\begin{table*}[ht]
\centering
\ra{1.3}
\begin{tabular}{lrrrrr}
\rowcolor{lightgray}
\vspace{-0.1cm}
&\multicolumn{5}{c}{PERCENTILES OF POSTERIOR}\\
\rowcolor{lightgray}
CAUSAL EFFECT OF 1SD CHANGE& \multicolumn{5}{c}{DISTRIBUTION}\\
\rowcolor{lightgray}
IN PROTEIN LEVEL ON MS RISK& $5$ & $25$ & $50$ & $75$ & $95$\\
\midrule
Causal Exposure Log Odds Ratio & -1.05 & -0.69 & -0.43& -0.19 & 0.14\\ 
  Causal Exposure Odds Ratio  & 0.35 & 0.50 & 0.65 & 0.82& 1.15  \\
\end{tabular}
\caption{\footnotesize Estimated causal effect of IL12A protein level on MS,
                 expressed on both a log-odds ratio and an odds ratio
                 scale, as obtained by an analysis that incorporates
                 both kinship information and the family indicator.}
\end{table*}

\newpar A comparison with the preceding table shows that introduction of the family variable left the point estimate of the causal effect substantially unchanged, while widening the credible interval, with a consequent, further, reduction in statistical significance of the result. This is hardly suprising, when one considers that families 3 and 7 (out of our 12 families) impacted on both exposure and outcome with effects of the same sign, as described later in this section. This will inevitably inflate association between exposure and outcome beyond the  component of association due to a genuinely causal relationship.

\vspace{1.6cm}
\begin{center}
{\Large Results from Final Model} 
\end{center}

\noindent In addition to kinship and to the 
family indicator, our final model includes the 
measured parental levels of circulating IL12A protein,
which means 
the protein level in the mother and in the father.
See Methods section for technical details.  This 
final elaboration increased the amount of instrumental 
information in the model, and produced the estimates 
summarized in Table 4. The point estimate for the causal 
effect of IL12A protein level on risk of MS was -0.49 on 
a log-odds ratio scale, and 0.61 on an odds-ratio scale. 
The corresponding 95$\%$ credible interval, also reported 
in Table 4, was entirely contained in the negative real axis, 
and included effect values of biological importance.

\vspace{1cm}
\begin{table*}[ht]
\centering
\ra{1.3}
\begin{tabular}{lrrrrr}
\rowcolor{lightgray}
\vspace{-0.1cm}
&\multicolumn{5}{c}{PERCENTILES OF POSTERIOR}\\
\rowcolor{lightgray}
CAUSAL EFFECT OF 1SD CHANGE& \multicolumn{5}{c}{DISTRIBUTION}\\
\rowcolor{lightgray}
IN PROTEIN LEVEL ON MS RISK& $5$ & $25$ & $50$ & $75$ & $95$\\
\toprule\\
Causal Log Odds Ratio Effect of Exposure on Outcome& -1.12 & -0.71 & -0.49 & -0.29 & -0.1 \\ 
  Causal Odds Ratio Effect of Exposure on Outcome  & 0.33 & 0.49 & 0.61 & 0.75 & 0.90 \\ 
\end{tabular}
\caption{\footnotesize Causal effect estimates from a model that incorporates kinship information, family indicator, and parental protein levels.}
\end{table*}
\vspace{1cm}

\newcommand{\midheader}[2]{%
\topmidheader{#1}{#2}}
\newcommand\topmidheader[2]{\multicolumn{#1}{c}{\textsc{#2}}\\%
                \addlinespace[0.5ex]}

\begin{table*}[ht]
\centering
\ra{1.3}
\begin{tabular}{lrrrrr}
\rowcolor{lightgray}
\vspace{-0.1cm}
& \multicolumn{5}{c}{PERCENTILES OF POSTERIOR}\\
\rowcolor{lightgray}
FAMILY& \multicolumn{5}{c}{DISTRIBUTION OF EFFECT}\\
\rowcolor{lightgray}
& $5$ & $25$ & $50$ & $75$ & $95$\\
\toprule\\
   \topmidheader{6}{\begin{minipage}{6.9cm}
Family-specific {\textbf {Indirect}} 
Causal Effect on Risk of MS (Odds Ratio)\end{minipage}}
  family 2  & 0.77 & 0.92 & 0.99 & 1.04 & 1.18  \\ 
  family 3  & 0.70 & 0.86 & 0.95 & 1.00 & 1.12  \\ 
  family 4 & 0.83 & 0.97 & 1.03 & 1.15 & 1.44 \\ 
  family 5 &  0.76 & 0.91 & 0.99 & 1.03 & 1.19  \\ 
  family 6 &  0.82 & 0.97 & 1.02 & 1.11 & 1.39  \\ 
  family 7 &  0.31 & 0.53 & 0.70 & 0.88 & 1.12  \\ 
  family 8 &  0.85 & 0.96 & 1.01 & 1.08 & 1.30 \\ 
  family 9 &  0.79 & 0.94 & 1.00 & 1.05 & 1.21 \\ 
  family 10 &  0.82 & 0.96 & 1.02 & 1.11 & 1.35  \\ 
  family 11 &  0.71 & 0.88 & 0.96 & 1.01 & 1.17 \\ 
  family 12 &  0.76 & 0.91 & 0.98 & 1.03 & 1.21 \\
  \midheader{6}{\vspace{0.3cm}\begin{minipage}{6.9cm} \vspace{0.4cm}
Family-specific {\textbf{Direct}} Causal Effect on Risk of MS (Odds Ratio)\vspace{0.2cm}\end{minipage}\vspace{-0.3cm}}
  family 2  & 0.20 & 0.44 & 0.73 & 1.09 & 1.74 \\ 
  family 3 & 0.47 & 0.86 & 1.28 & 1.90 & 3.33 \\ 
  family 4  & 0.56 & 1.09 & 1.82 & 3.43 & 10.04 \\ 
  family 5  & 0.33 & 0.66 & 1.00 & 1.49 & 2.70 \\ 
  family 6  & 0.26 & 0.58 & 0.94 & 1.44 & 2.84 \\ 
  family 7  & 0.53 & 0.97 & 1.58 & 2.65 & 5.67  \\ 
  family 8  & 0.36 & 0.69 & 1.07 & 1.61 & 3.09  \\ 
  family 9  & 0.27 & 0.54 & 0.81 & 1.21 & 2.23 \\ 
  family 10 & 0.16 & 0.42 & 0.67 & 1.05 & 1.86  \\ 
  family 11 & 0.58 & 1.04 & 1.59 & 2.39 & 4.28 \\ 
  family 12  & 0.67 & 1.15 & 1.64 & 2.38 & 4.24  \\ 
\end{tabular}
\caption{\footnotesize Additional results from our final model. For
each of the 12 families represented in our data, this table reports
the estimated effect that being a member
of that family has on MS risk, by distinguishing
between the direct and the indirect ($=$mediated by changes in the level
of circulating IL12A protein) components of the effect.
See rigorous definition of
direct and indirect effect in the
Methods section.}
\end{table*}

\newpar Figures 3 through 5 
summarize extra output of the analysis via our final model. These figures have been obtained by using the excellent {\tt bayesplot} package, writen in {\tt R} language by Jonah Gabry and colleagues \cite{Gabry2019}, as an aid to studying the output of {\tt Stan} analyses.

\newpar Figure 3
shows posterior intervals of for the instrument-exposure associations, $\alpha$. It is apparent from the figure that a few instruments, eg. instrument 49, stand out in terms of strength. The sparsity prior we have imposed on these effects is able to pick up the few "needles in the haystack", while downplaying the role of weaker instruments, at the same time working in the direction of a reduction of the weak instrument bias. It might be interesting to investigate the strong instruments from a functional point of view.

\newpar Figure 4
shows posterior intervals for familial effects on outcome, that we call "direct" because they are not mediated by the exposure. One may wish to interpret these as familial effects mediated by IL12A-independent pathways, environment and lifestyle. The figure highlights some families (eg., family 12) as characterized by a higher risk of MS, compared with the others. In a separate work we investigate the factors responsible of such differences in detail. Other families (eg., family 2) appear to "protected" from MS due to factors other than IL12A.

\newpar Figure 5
shows posterior intervals for familial effects on outcome, that we call "indirect" because they are mediated by the exposure. They are calculated  by including in the model a parameter defined to represent the product of the $F \rightarrow X$ effect and the $X \rightarrow Y$ effect. The posterior distribution for this parameter gets sampled by the MCMC inference engine. The sample are then automatically used to calculate posterior mean and credible interval.  A comparison between Figures 4 and 5 suggests that in certain families, eg. family 4 in our sample, both the direct and the indirect effects operate deleteriously, whereas in others, eg. family 7, the two effects tend to cancel each other.

\begin{figure}[!h]
\label{Figure 3}
\caption{\footnotesize Estimated posterior intervals for the instrument-exposure associations, $\alpha$, based on our analysis with the final, complete, Bayesian model, that includes kinship information, family variable and parental protein levels.}
\vspace{-0.4cm}
\begin{center}
\scalebox{1.4}{
\includegraphics[width=0.45\textwidth]{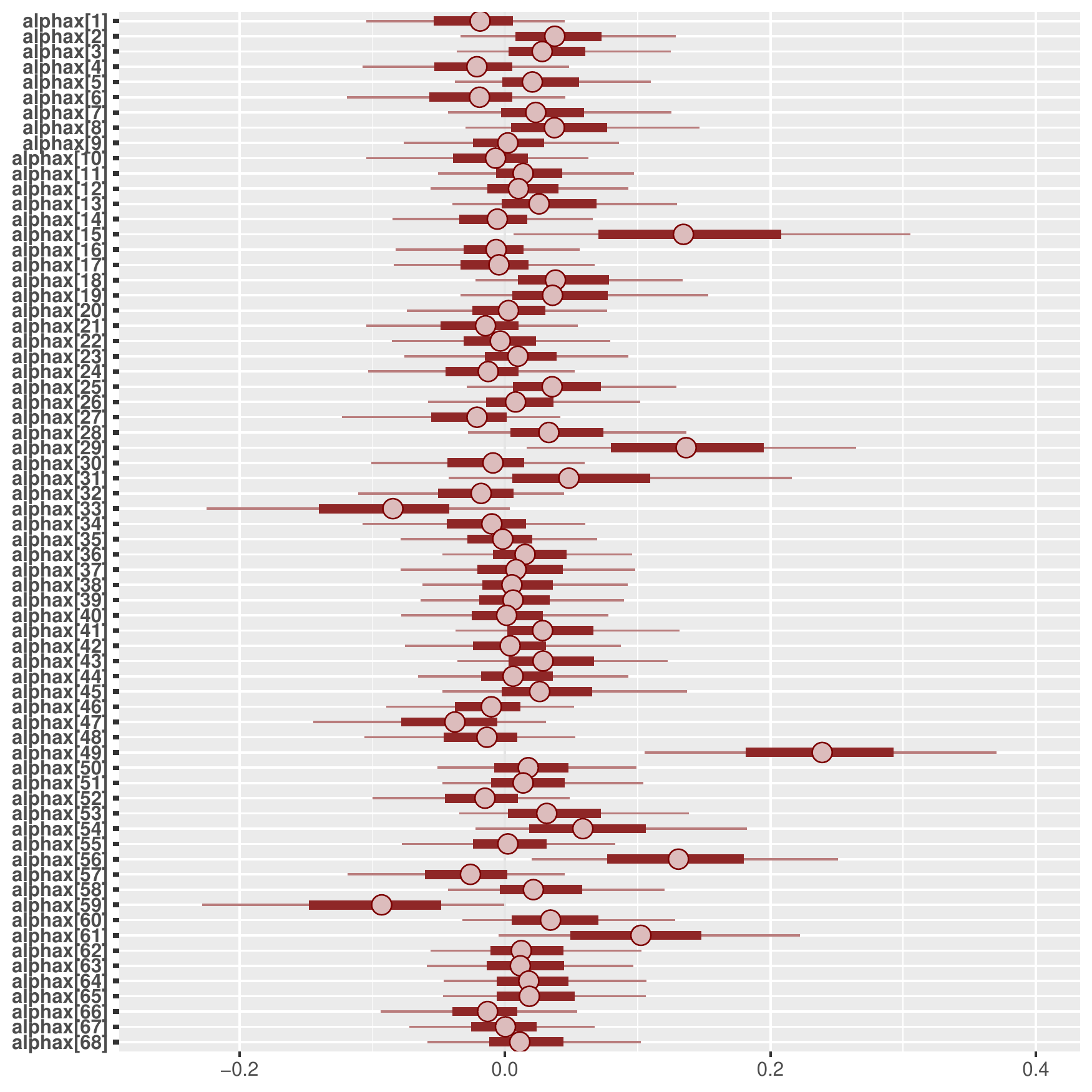}
}
\end{center}
\end{figure}

\begin{figure}[!h]
\label{Figure 4}
\vspace{-0.4cm}
\begin{center}
\scalebox{1.4}{
\includegraphics[width=0.45\textwidth]{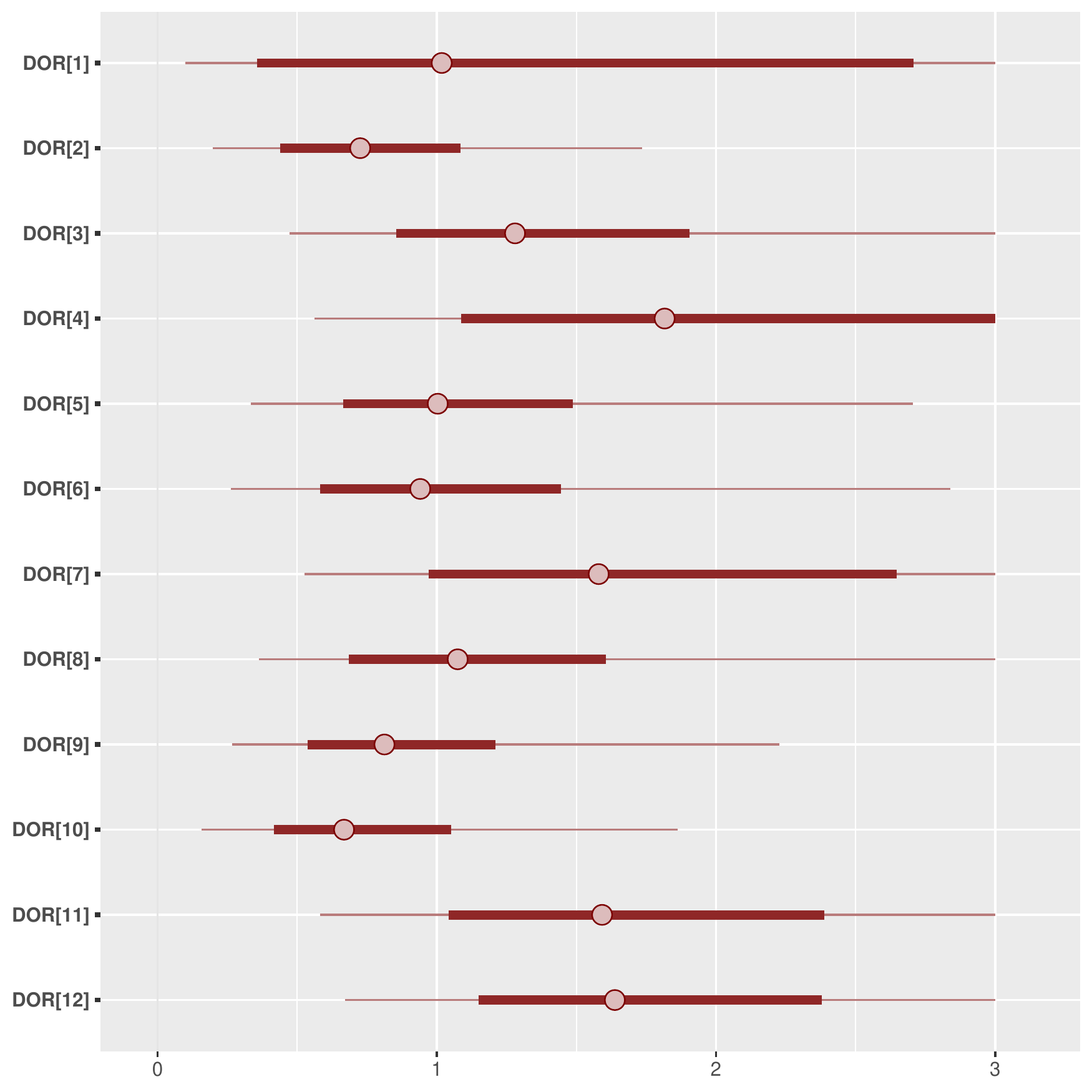}
}
\end{center}
\caption{\footnotesize Estimated direct causal effects of family membership on risk of MS, expressed on an odds-ratio scale, based on our analysis with the final, complete, Bayesian model, that includes kinship information, family indicator and parental protein levels.}
\end{figure}

\begin{figure}[!h]
\label{Figure 5}
\vspace{-0.4cm}
\begin{center}
\scalebox{1.4}{
\includegraphics[width=0.45\textwidth]{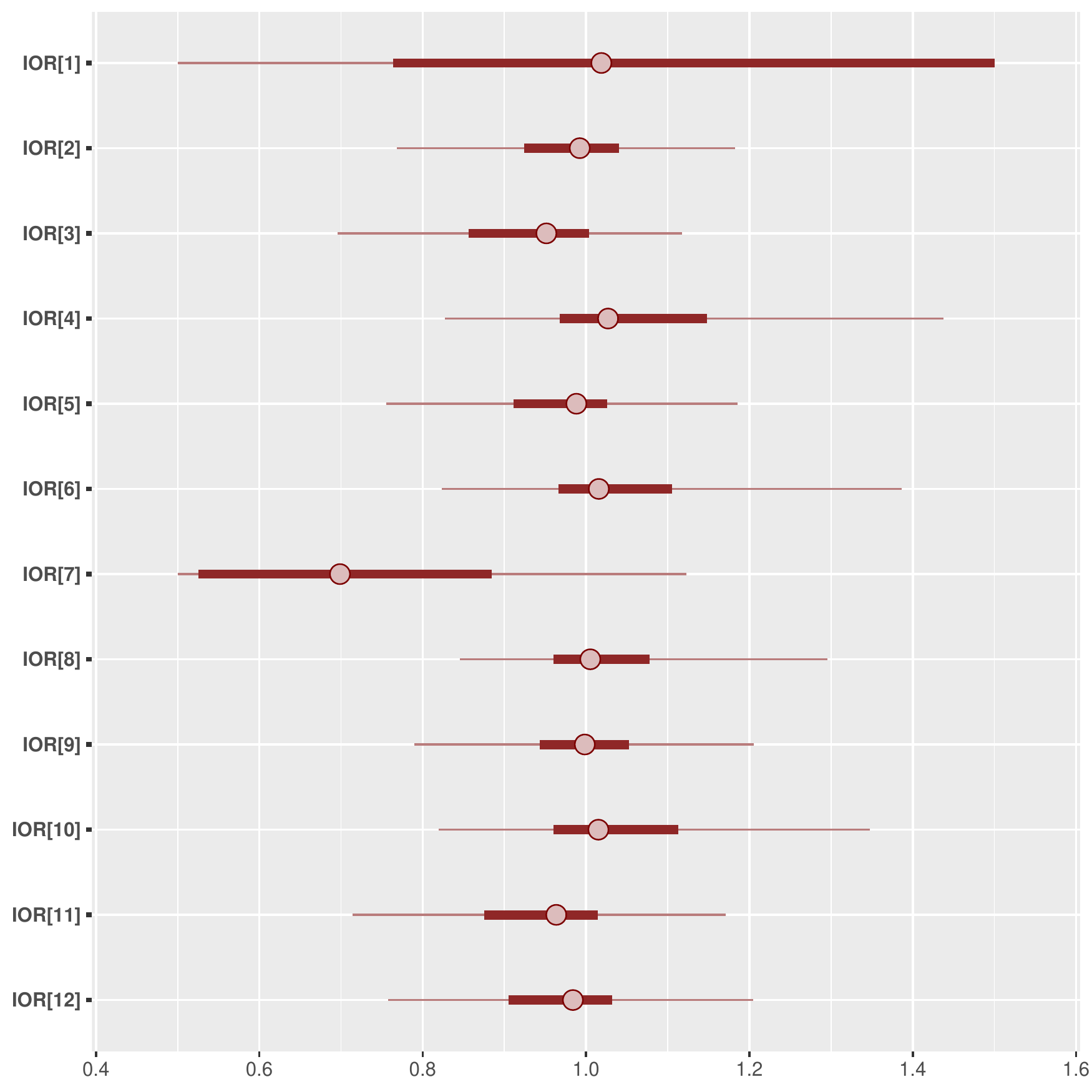}
}
\end{center}
\caption{\footnotesize Estimated indirect causal effects of family membership on risk of MS, expressed on an odds-ratio scale, based on our analysis with the final, complete, Bayesian model, that includes kinship information, family indicator and parental protein levels. We use the term "indirect" to signify the effect on MS risk exerted by membership to a particular family through the mediation of IL12A plasma level.}
\end{figure}

\newpar We calculated posterior predictive check diagnostics based on discrepancies between (a continuous approximation of) the observed outcome variable distribution and the corresponding distribution generated from the posterior values of the unknown parameters of this final model. No signal of model misfit has been found (see Supplementary Material).

\begin{figure}[!h]
\label{Pathway}
\vspace{0.4cm}
\begin{center}
\scalebox{1.8}{
\includegraphics[width=0.45\textwidth]{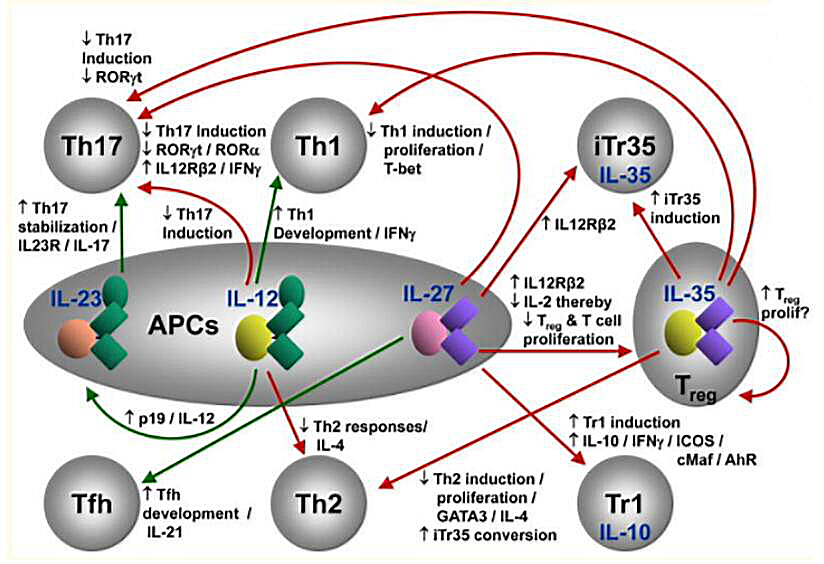}
}
\end{center}
\caption{\footnotesize IL12 family cytokines as a
putative immunological link between IL12-A and MS.}
\end{figure}

\newpar Gene IL12A (p35), together with gene IL12B (p40), encodes 
Interleukin 12 (abbreviated: IL12). IL12 is a pro-inflammatory 
cytokine, produced mainly by antigen presenting 
cells (abbreviated: APCs). It acts as an immunological 
playmaker by inducing Th1 cell differentiation from CD4$+$ 
naive T cells, interferon $\gamma$ (abbreviated: IFN-$\gamma$) 
production and tumor necrosis factor-alpha (abbreviated: 
TNF-$\alpha$) from T cells and natural killer (abbreviated: NK) 
cells \cite{Aslani2017}. A diagrammatic picture of the relevant 
pathway is shown in Figure 6.
The hypothesised causal effect of IL12A on risk of 
MS might be mediated by the encoding of IL12 and the 
subsequent IL12-induced production of IFN-$\gamma$. 
In fact, IFN-$\gamma$ is a major cytokine found in MS 
lesions, and it has been found that its levels are greatly 
increased during MS activity \cite{Lees2007}. IL12-induced 
IFN-$\gamma$ production is the key point in the Th1 immune 
responses induction and proliferation.

\newpar Furthermore, in murine models, IL12 has been shown to 
induce Substance P (SP) precursor mRNA in macrophages via STAT4 
pathway \cite{Arsenescu2005} and NK1R expression by both IL12 and 
IL18 stimulation via NF$\kappa$B in T cells \cite{Weinstock2003}. 
SP has a demonstrated role in neuroimmune, autoimmune and 
inflammatory conditions, including MS \cite{Oconnor2004, 
Kostyk1989}. But while IL12 and IL23 are pro-inflammatory 
cytokines, IL27 and IL35 are inhibitory cytokines. So, clearly, 
their immune balance is crucial for the modulation of immune function.

\section*{\Large Discussion}

\noindent We have extended the Bayesian MR framework of Berzuini and 
colleagues \cite{Berzuini2018a} for use in the analysis of pedigree data. 
MR has only rarely been applied to this class of data. Also, MR has been 
most frequently applied to the study of high-level exposures, such as 
as obesity \cite{Conde2018,Mariosa2019}, whereas our illustrative 
application deals with a molecular exposure. Some researchers appear 
confident that standard MR methods work equally well with molecular 
exposures, such as transcripts and proteins. Our early experiences in 
this area do not entirely corroborate this optimism, one reason being 
the intrinsic paucity of instruments at a molecular level. Although public 
bioinformatic repositories are sprawling with data, the number of available 
instruments for the analysis of causality at a molecular level is generally, 
and inevitably, poor due the the intrinsic nature of the studied mechanism. 
This makes MR analyses extremely vulnerable to the presence of confounding, 
not least because of possible, untestable, violations of the confounder 
independence assumption. MR analysis of pedigree data (as opposed to 
samples of unrelateds) promises robustness to confounding, and, for 
this reason, it presents itself as a useful tool for dealing with the 
information weakness we encounter in the study of causality at a 
molecular level.  Motivated by these considerations, we have extended 
MR to work with pedigree data.

\newpar Results of our illustrative study point to the circulating 
level of protein IL12A as a potential cause of MS. While unexciting 
from a statistical significance viewpoint, our results match existing 
biological evidence. Interleukin 12 (IL12) is a pro-inflammatory cytokine, produced mainly by Antigen Presenting Cells (APCs). 
IL12 is a heterodimeric cytokine encoded by two separate genes, IL-12A (p35) and IL-12B (p40). 
It acts as an immunological playmaker inducing Th1 cell differentiation from CD4+ naive T cells, interferon $\gamma$ (IFN-$\gamma$) production and tumor necrosis factor-alpha (TNF-$\alpha$) from T cells and natural killer (NK) cells \cite{Aslani2017, Katan1986}. 
IFN-$\gamma$ is a major cytokine found in MS lesions, and its levels are greatly increased during MS activity \cite{lees2007little}.
IFN-$\gamma$ production induced by IL12 is the key point in the Th1 immune responses induction and proliferation.
Furthermore, in murine models, IL-12 has been shown to induce precursor mRNA of Substance P (SP) in macrophages via STAT4 pathway (IL-12 induction of mRNA encoding substance P in murine macrophages from the spleen and sites of inflammation \cite{Arsenescu2005}). In addition, both IL-12 and IL-18 stimulation induces NK1R expression via NF$k$B pathway in T cells (IL-18 and IL-12 signal through the NF-kappa B pathway to induce NK-1R expression on T cells \cite{Weinstock2003}). SP has a demonstrated role in neuroimmune, autoimmune and inflammatory conditions, including MS \cite{Oconnor2004} \cite{Kostyk1989}.
As shown in the figure 6, while IL-12 and IL-23 are pro-inflammatory cytokines, on the contrary IL-27 and IL-35 are inhibitory cytokines. So clearly, the immune balance of all the cytokines involved  is crucial for the modulation of immune function where compensatory  mechanisms can play a strategic role, that may explain the negative sign of the causal effect we found that is in contradiction with the expected increase of MS risk induced by IL12A. 

\newpar From a statistical viewpoint, our IL12A data analysis illustrates a few important points. Firstly, because introduction of kinship information in the model accounts for the reduction in the number of "effective" individuals due to family correlation, it may result in an increased posterior uncertainty about the causal effect, with a reduction of evidence against the null causal hypothesis. Introduction of the family indicator may have a similar effect on the causal estimate, that of a greater posterior uncertainty, with a consequent further reduction of evidence of causality. Recall that family membership is a potential instrument-outcome confounder. The increase in posterior uncertainty consequent to introduction of the family indicator may thus be interpreted as an effect of the de-biasing. Our results suggest that our elaborations of the models tend to avoid over-optimistic results, which we believe to work in the direction of a healtier science. Parental protein information, introduced at the last model elaboration step, acted as instrumental, which resulted in an increase of evidence of causality.

\newpar MR has been traditionally applied to data from unrelated 
individuals. This is a pity, because MR analysis of family data is 
inherently more robust to population stratification and heterogeneity 
than analysis of untelateds. We believe this property  to help 
disentangle inheritable from environmental effects. A potentially 
fruitful idea is to collect data from unrelated individuals and 
then to collect further data from the parents of those individuals, 
for a joint analysis of the two data sources. Such a joint analysis 
can be performed via our proposed approach by treating parent-child 
triads as "families". Or one could use information from previous 
analyses of unrelateds in order to shape informative priors for an 
analysis of pedigree data along our proposed lines. Pedigree 
analysis might prove an invaluable tool for studying disease 
mechanism peculiarities of small, possibly native and isolated, 
populations. We are, in particular, thinking of small populations 
characterized by maverick disease patterns, that suffer from 
inadequate attention from the medical research community, 
perhaps outside the western "white" world.

\newpar Finally, on a more methodological note, we would emphasize 
the flexibility of a MCMC-powered Bayesian approach in MR, especially 
thanks to the possibility of straightforward elaboration of the basic 
MR model to accommodate extra relevant information and the 
straightforward handling of missing information.

\newpar We are at present working on an extension of the models discussed here to incorporate haplotype information.

\bibliographystyle{plain}
\bibliography{bibliografia}

\newpage

\section*{Stan code}
\noindent \verb&BayesianMR <-'&\\
\x data {&\\
\x int<lower=0> nobs;&\hfill number of individuals with non-missing value of $X$
\x int<lower=0> nmis;&\hfill number of individuals with missing value of $X$\\
\x int<lower=0> nfam;&\hfill total number of families\\
\x int<lower=0> N;&\hfill total number of sample individuals ($=$NOBS+NMIS)\\
\x int<lower=0> J;&\hfill total number of instruments\\
\x matrix[N,J] Z;&\hfill matrix of standardized (mean$=0$, SD$=1$) values of instruments\\
\x matrix[N,N] pedigree_matrix;&\hfill kinship matrix over whole sample\\
\x vector[nobs] Xobs;&\hfill observed $X$ values, transformed to a $0$-mean-$1$-SD variables\\
\x int Y[N];&\hfill vector of (0,1) disease indicators\\
\x vector[N] PROTEINMADRE;&\hfill measured protein level in mother\\
\x vector[N] PROTEINPADRE;&\hfill measured protein level in father\\
\x real<lower=0> betasimsd;&\\
\x real<lower=0> cauchysd;&\hfill scale parameter for Cauchy prior on causal parameter\\
\x real<lower=1> nu_global;&\hfill degrees of freedom for the half-$t$ prior for tau\\
\x real<lower=1> nu_local;&\hfill df half-$t$ priors for the LAMBDAs (1$\rightarrow $horseshoe)\\
\x matrix[N,nfam] FAM;&\hfill matrix of standardized (mean$=0$,SD$=1$) family indicators\\
\x vector[N] muY;&\hfill auxiliary\\
\x }&\\
\\
\newpar\verb&transformed data{&\\
\x matrix[N,N] L_pedigree_matrix = cholesky_decompose(pedigree_matrix);&\\
\x }&\\
\\
\newpar\verb&parameters {&\\
\x real<lower=0,upper=1> frazionepleio;&\hfill expected fraction pleiotrop. instr.\\
\x real <lower=0> sigmax;&\hfill lik-unidentifiable SD of measurement error on $X$\\
\x real <lower=0> sigmaalpha;&\hfill SD of ALPHAX hyperprior\\
\x real theta;&\hfill causal effect of inferential interest\\
\x real <lower=-1,upper=1> alphaMADRE;&\\
\x real <lower=-1,upper=1> alphaPADRE;&\\
\x real omegay;&\hfill intercept of the model for $Y$\\
\x real deltax;&\hfill effect of $U$ on $X$, and covariance betw $X$- and $Y$-errors\\
\x vector[N] u;&\hfill individual-specific confounder values\\
\x vector[J] alphax;&\hfill effects of instruments on exposure\\
\x vector[nmis] Xmis;&\hfill unobserved subset of values of $X$\\
\x vector[nfam] gammafamx;&\\
\x vector[nfam] gammafamy;&\\
\x vector[N] correction;&\\
\\

\newpar \hfill Auxiliary variables that define the global and local parameters:\\
\x vector[J] z;&\\
\x real<lower=0> r1_global;&\\
\x real<lower=0> r2_global;&\\
\x vector<lower=0>[J] r1_local;&\\
\x vector<lower=0>[J] r2_local;&\\
\x }&\\
\\

\newpar \verb&transformed parameters {&\\

\hfill Half-$t$ priors for the lambdas:\\
\x vector[J] beta;&\hfill unknown pleiotropic effects in real dataset\\
\x real<lower=0> tau;&\hfill global shrinkage parameter\\
\x vector<lower=0>[J] lambda;&\hfill local shrinkage parameter\\
\x real<lower=0> m0;&\\
\x real<lower=0> scale_global;&\\
\x vector[N] Xcompleto;&\\
\x lambda = r1_local .* sqrt(r2_local);&\\
\x tau= r1_global * sqrt(r2_global);&\\
\x beta = z .* lambda*tau;&\\
\x m0=floor(J*frazionepleio);&\hfill expected number of large pleiotropic effects\\
\x scale_global= 2*m0/(sqrt(N)*(J-m0));&\\
\x Xcompleto[1:nobs]= Xobs;&\\
\x Xcompleto[(nobs+1):N] = Xmis;&\\
\x }&\\
\\

\newpar\verb&model {&\\
\x frazionepleio ~ uniform(0.1,0.9);&\\
\x z~ normal(0,1);&\\
\x r1_local~ normal(0.0,1.0);&\\
\x r2_local~ inv_gamma(0.5*nu_local,0.5*nu_local);&\\
\x r1_global~ normal(0.0,scale_global);&\hfill required for half-$t$ prior for TAU\\
\x r2_global~ inv_gamma(0.5*nu_global,0.5*nu_global);&\hfill as above\\
\\

\newpar\hfill Model for observed values of $X$\\
\x Xobs     ~ normal(&\\                   
\hspace*{1cm}\x FAM[1:nobs,]*gammafamx&\hfill family $\rightarrow X$\\
\hspace*{1cm}\x +Z[1:nobs,]*alphax&\hfill instruments $\rightarrow X$\\
\hspace*{1cm}\x +PROTEINMADRE[1:nobs]*alphaMADRE&\hfill maternal protein level $\rightarrow X$\\
\hspace*{1cm}\x +PROTEINPADRE[1:nobs]*alphaPADRE&\\
\hspace*{1cm}\x +u[1:nobs]*deltax,&\hfill unknown confounder $\rightarrow X$\\
\x sigmax);&\\
\\

\newpar\hfill Model for unobserved values of X,\\
\x Xmis~ normal(&\hfill to be imputed as part of inference\\
\hspace*{1cm}\x FAM[(nobs+1):N,]*gammafamx+Z[(nobs+1):N,]*alphax&\\
\hspace*{1cm}\x +PROTEINMADRE[(nobs+1):N]*alphaMADRE&\\
\hspace*{1cm}\x +PROTEINPADRE[(nobs+1):N]*alphaPADRE&\\
\hspace*{1cm}\x +u[(nobs+1):N]*deltax,&\\
\x sigmax);&\\
\\

\newpar\hfill Observation model for $Y$\\
\x correction ~ multi_normal_cholesky(muY, L_pedigree_matrix);&\\
\x Y ~ bernoulli_logit(omegay+Z*beta +FAM*gammafamy&\\
\hspace*{1cm}\x +Xcompleto*theta +u +correction);&\\
\\

\newpar\hfill Prior\\
\x theta ~ cauchy(0,cauchysd);&\\
\x for(n in 1:N){&\\
\hspace*{0cm}\x u[n] ~ normal(0,1);}&\\
\hspace*{0cm}\x alphaMADRE ~ uniform(-1,1);&\\
\x alphaPADRE ~ uniform(-1,1);&\\
\x for(h in 1:nfam){&\\
\hspace*{0.5cm}\x gammafamx[h] ~ cauchy(0,cauchysd);&\\
\hspace*{0.5cm}\x gammafamy[h] ~ cauchy(0,cauchysd);}&\\
\hspace*{0.5cm}\x for(k in 1:J){&\\
\hspace*{1cm}\x alphax[k]~ double_exponential(0, sigmaalpha);&\hfill note $0$ mean\\
\x }&\\
\x }&\\
'\\

\end{document}